\documentclass[3p,times,procedia]{elsarticle}
\usepackage{nupha_ecrc}

\volume{00}

\firstpage{1}

\journalname{Nuclear Physics A}

\runauth{}

\jid{nupha}

\jnltitlelogo{Nuclear Physics A}

\usepackage{amssymb}

\biboptions{sort&compress}

\usepackage[figuresright]{rotating}

\begin{document}

\begin{frontmatter}



\dochead{}

\title{Overview of recent results from the STAR experiment}


\author{Mustafa Mustafa (for the STAR Collaboration)}

\address{Lawrence Berkeley National Laboratory\\ MS70R0319, 1 Cycltron Road, Berkeley, CA 94720, USA}

\begin{abstract}
  The Solenoidal Tracker at RHIC (STAR) experiment utilizes its excellent mid-rapidity tracking and particle identification capabilities to study the emergent
properties of Quantum Chromodynamics (QCD). The STAR heavy-ion program at vanishingly small baryon density is aimed to address questions about the quantitative properties
of the strongly-interacting Quark Gluon Plasma (QGP) matter created in high energy collisions ($\eta/s$, $\hat{q}$, chirality, transport parameters, heavy quark diffusion coefficients ...).
At finite baryon density, the questions concern the phases of nuclear matter (the QCD phase diagram) and the nature of the phase transition, namely: what is the onset collision energy
for the formation of QGP? What is the nature of phase transition in heavy-ion collisions? Are there two phase transition regions?
If yes, where is the critical point situated? At Quark Matter 2015, the STAR collaboration has presented a wealth of new
experimental results which address these questions. In these proceedings I highlight a few of those results.

\end{abstract}


\end{frontmatter}


\section{Introduction}
\label{sec:introduction}
At Quark Matter 2015, STAR has presented results from its recently installed specialized detectors:
1) The Heavy Flavor Tracker (HFT), which facilitates the topological reconstruction of heavy flavor hadrons~\cite{giacomo} and
 2) The Muon Telescope Detector (MTD) which enables the identification of muons and consequently
 the study of quarkonia via the di-muon channel. The HFT has enabled, for the first time, the measurement of $D^0$ azimuthal anisotropy
 at RHIC, a measurement that will help constrain the QGP transport coefficients. The first quarknonium measurements via the di-muon channel using the MTD
 have also been presented. The RHIC Beam Energy Scan phase-I has recently been completed with Au+Au data taken at $\sqrt{s_{NN}} = 14.5$ GeV which filled the previous 100 MeV gap in baryon chemical
 potential (BES-I covers $20 < \mu_B < 420$ MeV). Directed, elliptic and triangular flow together with spectra, nuclear modification factors and higher-moments measurements from the full
 datasets of BES-I have been shown at this conference.

\section{Quantifying the properties of QGP}
\subsection{$D^{0}$ mesons measurements}

The large masses of heavy quarks (charm and bottom) endow them with unique properties that make them a
calibrated probe of the QGP medium created in heavy-ion collisions. Their masses are larger than all the relevant medium and QCD
scales ($m_{u,d,s}$, $\Lambda_{QCD}$, T), furthermore, their masses are mostly external to QCD, thus are not modified by their
presence in the medium in a scenario where the QCD condensate melts~\cite{Muller200584}. Their large masses also make
them amenable to Brownian motion theoretical-treatment in the QGP medium~\cite{rappHQReview}. $D^{0}$ $R_{AA}$ and $v_2$
are especially sensitive to the medium dynamics and thus have the potential to experimentally constrain the QGP transport coefficients~\cite{charmHTL}.

\begin{figure}[h!]
  \begin{center}
    \includegraphics[width=0.32\textwidth]{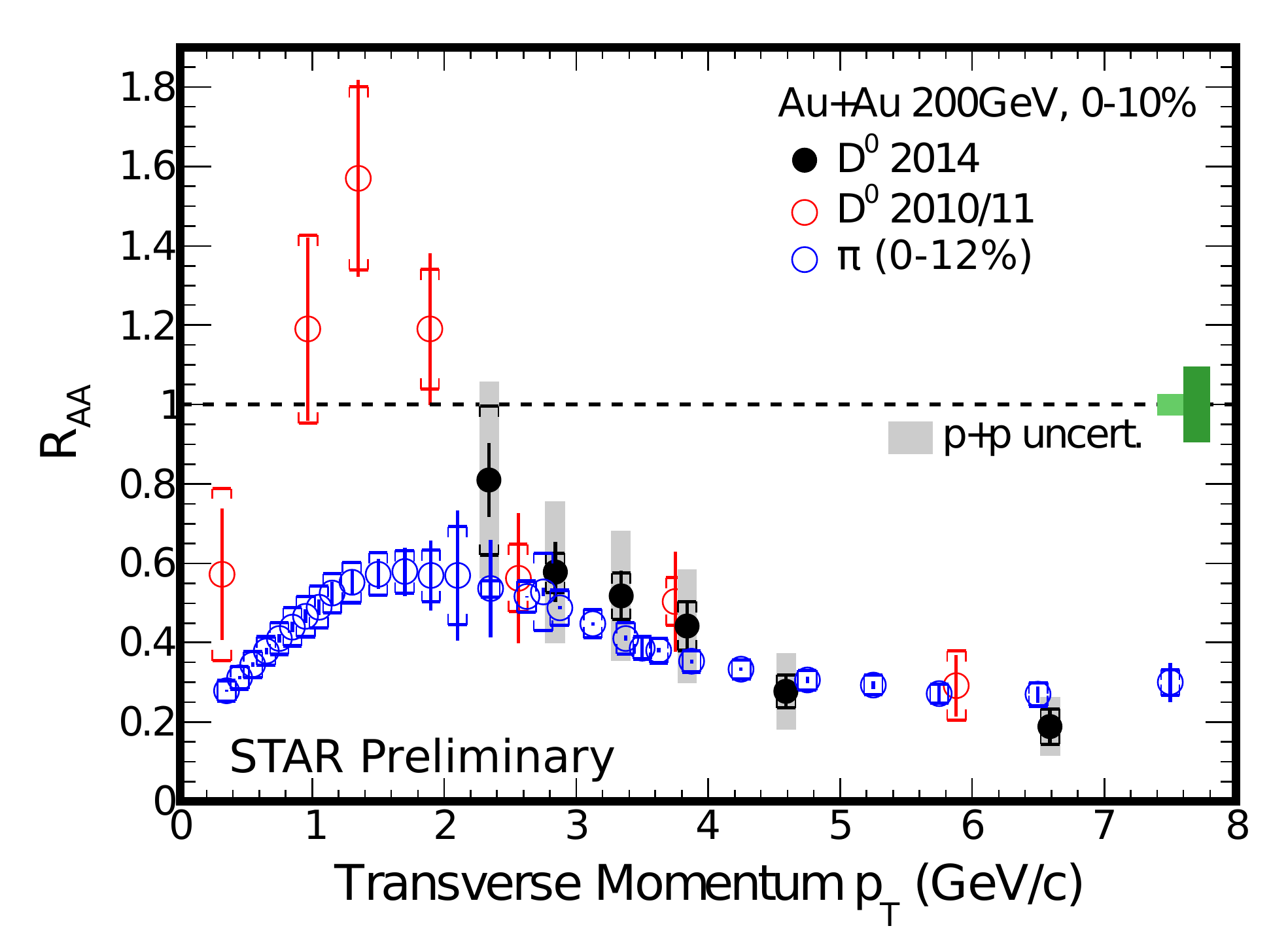}
    \includegraphics[width=0.32\textwidth]{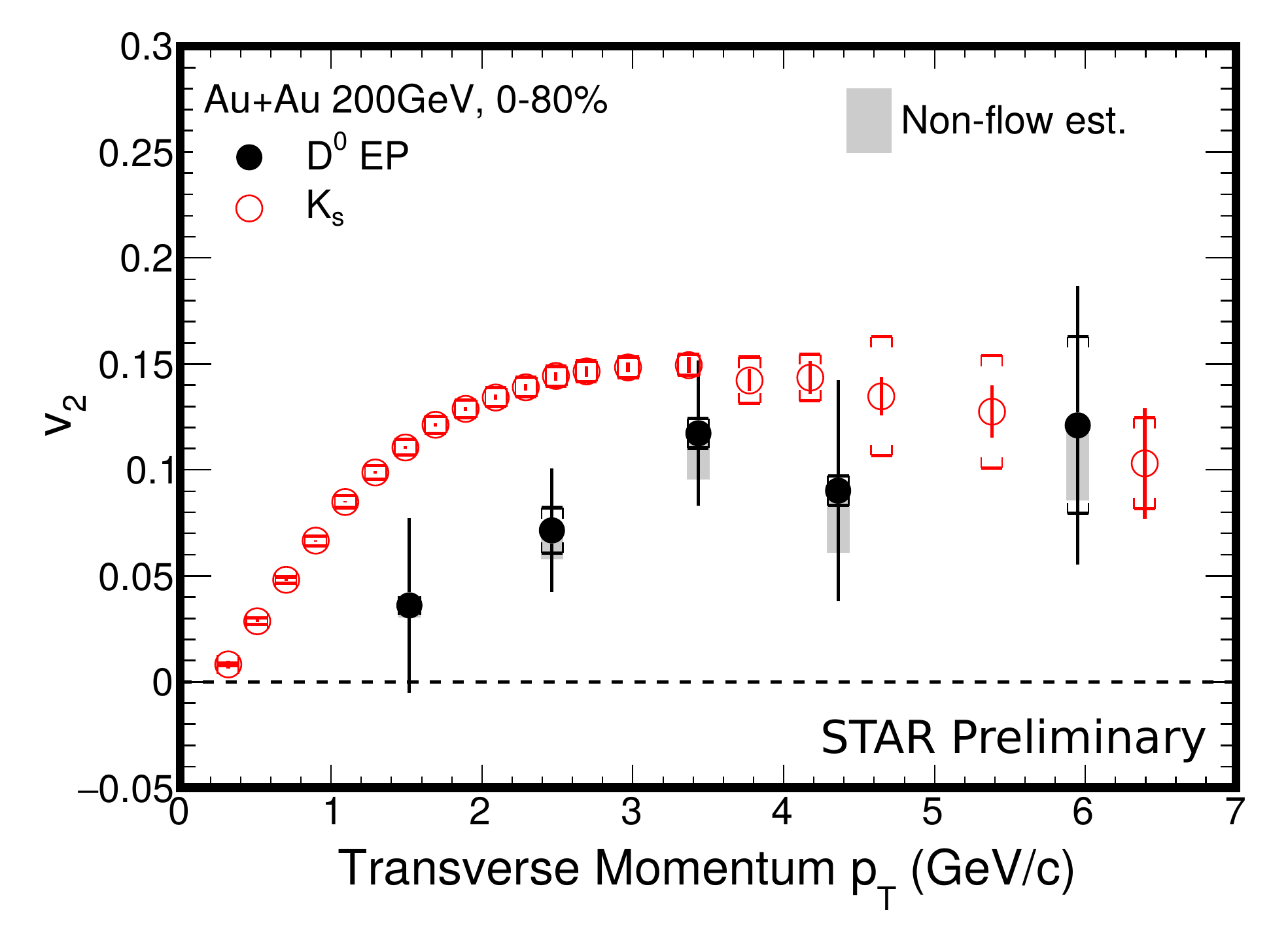}
    \includegraphics[width=0.32\textwidth]{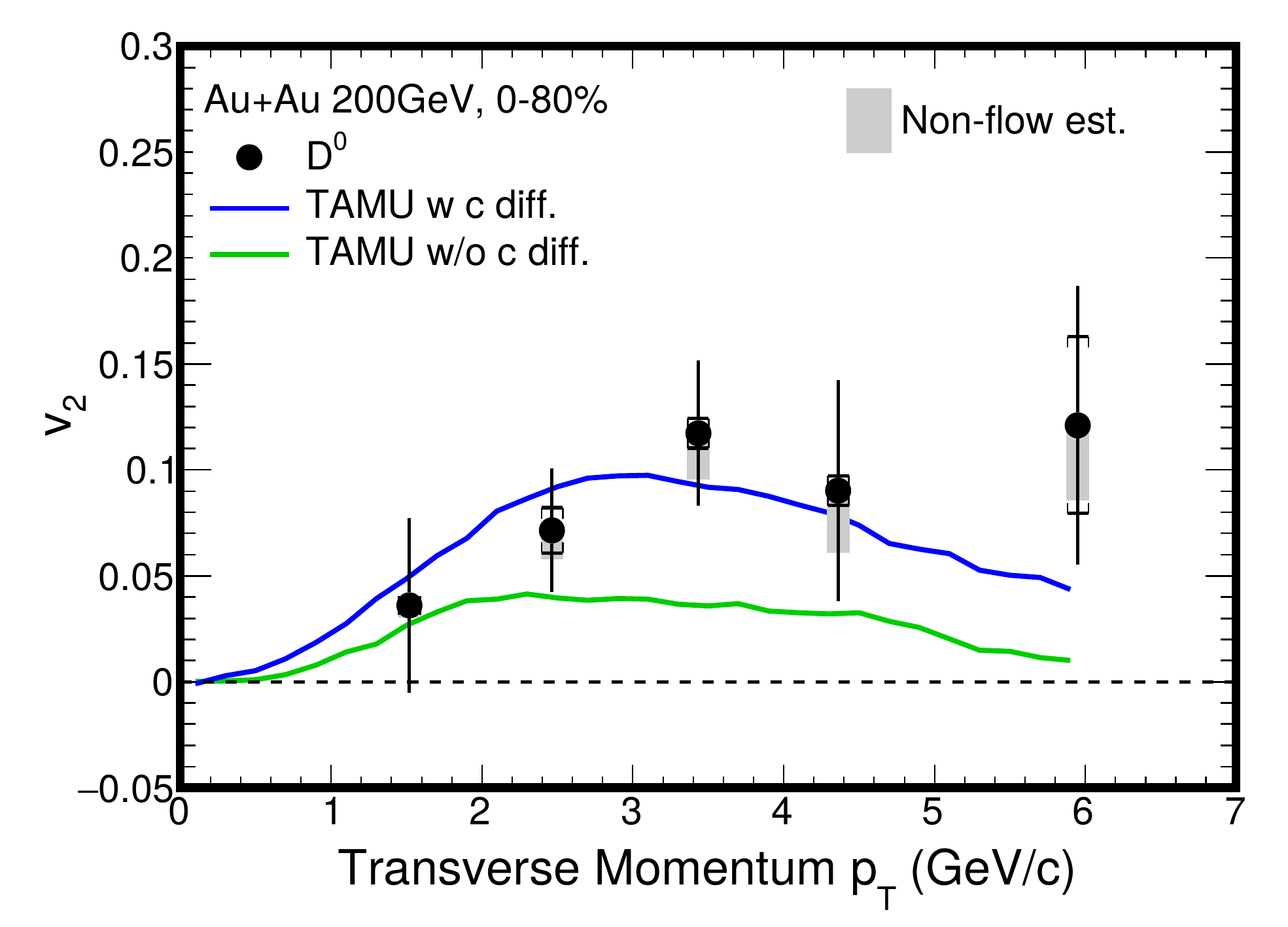}
    \caption{(Left) Nuclear modification factor of $D^{0}$ (red open symbols are previous results without HFT~\cite{publishedD0}, black filled symbols are
      the new measurements using HFT) and pions (blue open symbols)~\cite{pionsRaa} in central Au+Au collisions at $\sqrt{s_{NN}} = 200$ GeV.
    (Middle) $v_2$ of $D^{0}$ (filled symbols) compared to that of $K_s$ (open symbols)~\cite{ksV2} in minimum-bias collisions at $\sqrt{s_{NN}} = 200$GeV.
  (Right) $v_2$ of $D^{0}$ compared to TAMU model calculations with (blue) and without (green) charm diffusion~\cite{rapp}.}
  \label{fig:d0_data}
  \end{center}
\end{figure}

Figure \ref{fig:d0_data} left shows STAR measurements of $D^0$ $R_{AA}$~\cite{guannan}, the open red circles show the measurement
before the completion of the HFT~\cite{publishedD0}, solid black circles show the recent measurement with the HFT. The HFT clearly
helped reduce the uncertainties in Au+Au data especially in the transition region between the enhancement region at low-$p_T$
and the suppression region at high-$p_T$. The current uncertainties from the $p$+$p$ baseline are expected to be greatly reduced with the
STAR data collected in year 2015 with the HFT in place. The enhancement in the low-$p_T$ region is understood as a manifestation of charm
coalescence with a flowing medium while the suppression at high-$p_T$ is, according to model calculations, an interplay of elastic collisional and
radiative charm quark energy loss in the medium. While the suppression of $D^0$ is, within the current experimental uncertainties,
similar to that of light hadrons at high-$p_T$, radiative energy loss alone seems to be sufficient to explain the energy loss of light quarks~\cite{rappHQReview}.

Figure \ref{fig:d0_data} middle shows the first measurement of $D^0$ $v_2$ at RHIC top energy~\cite{michael}. Finite $v_2$ is observed for $p_T > 2$ GeV/c
that is systematically lower than those of lighter hadrons ($p_T < 4.0$ GeV/c). Figure \ref{fig:d0_data} right shows a comparison of
$D^0$ $v_2$ to TAMU calculation with and without charm diffusion~\cite{rapp}, the data are more compatible with charm diffusion in the medium,
 establishing a strong experimental evidence for charm quark exhibition of collectivity with the medium at RHIC top energy. The ordering, which is not explained by hydro mass ordering, might be hinting at
an incomplete thermalization of charm quarks with the medium. Higher precision data from RHIC Run 2016 will help us investigate this.

\begin{figure}[h!]
  \begin{center}
    \includegraphics[width=0.31\textwidth]{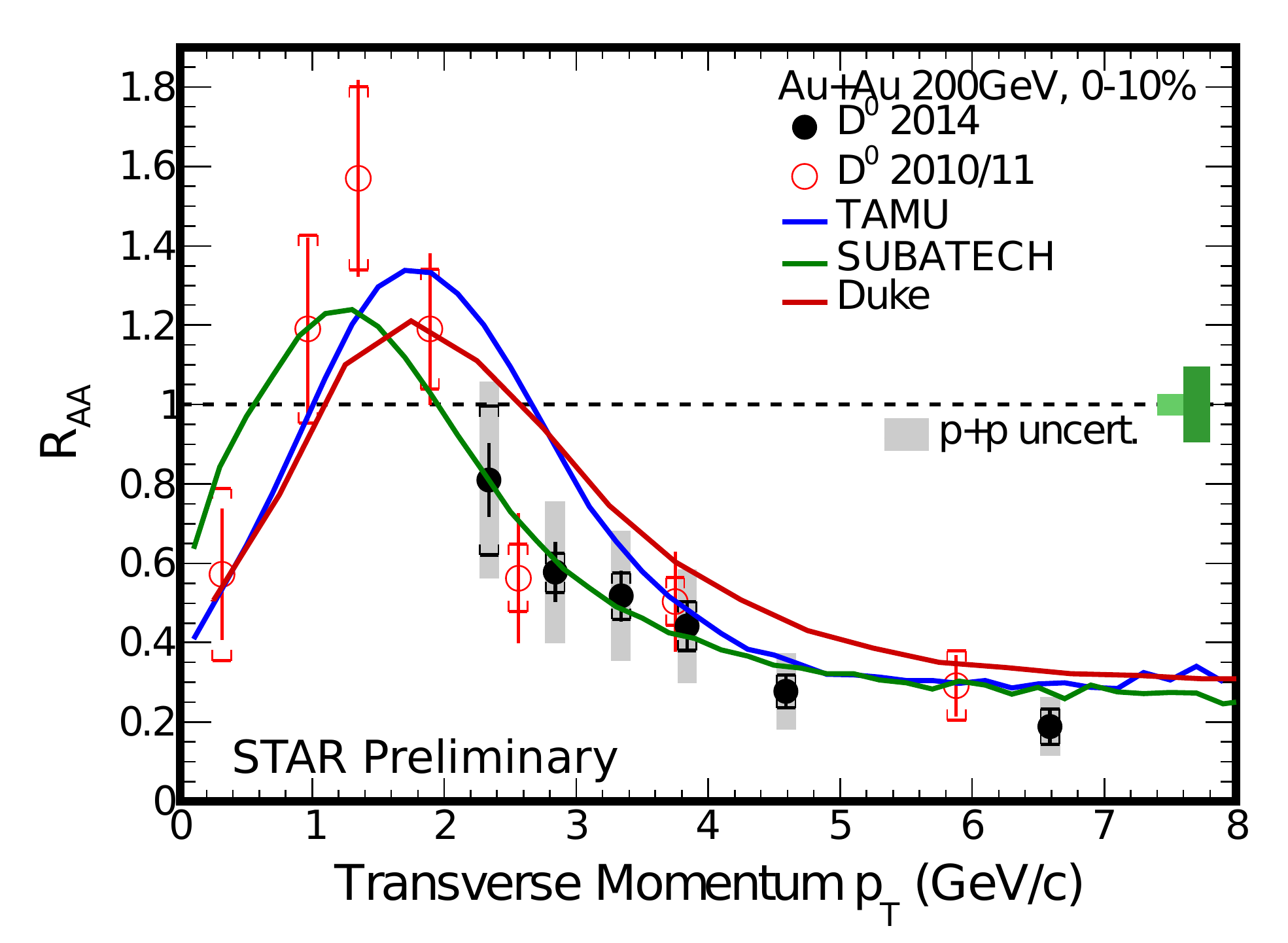}
    \includegraphics[width=0.31\textwidth]{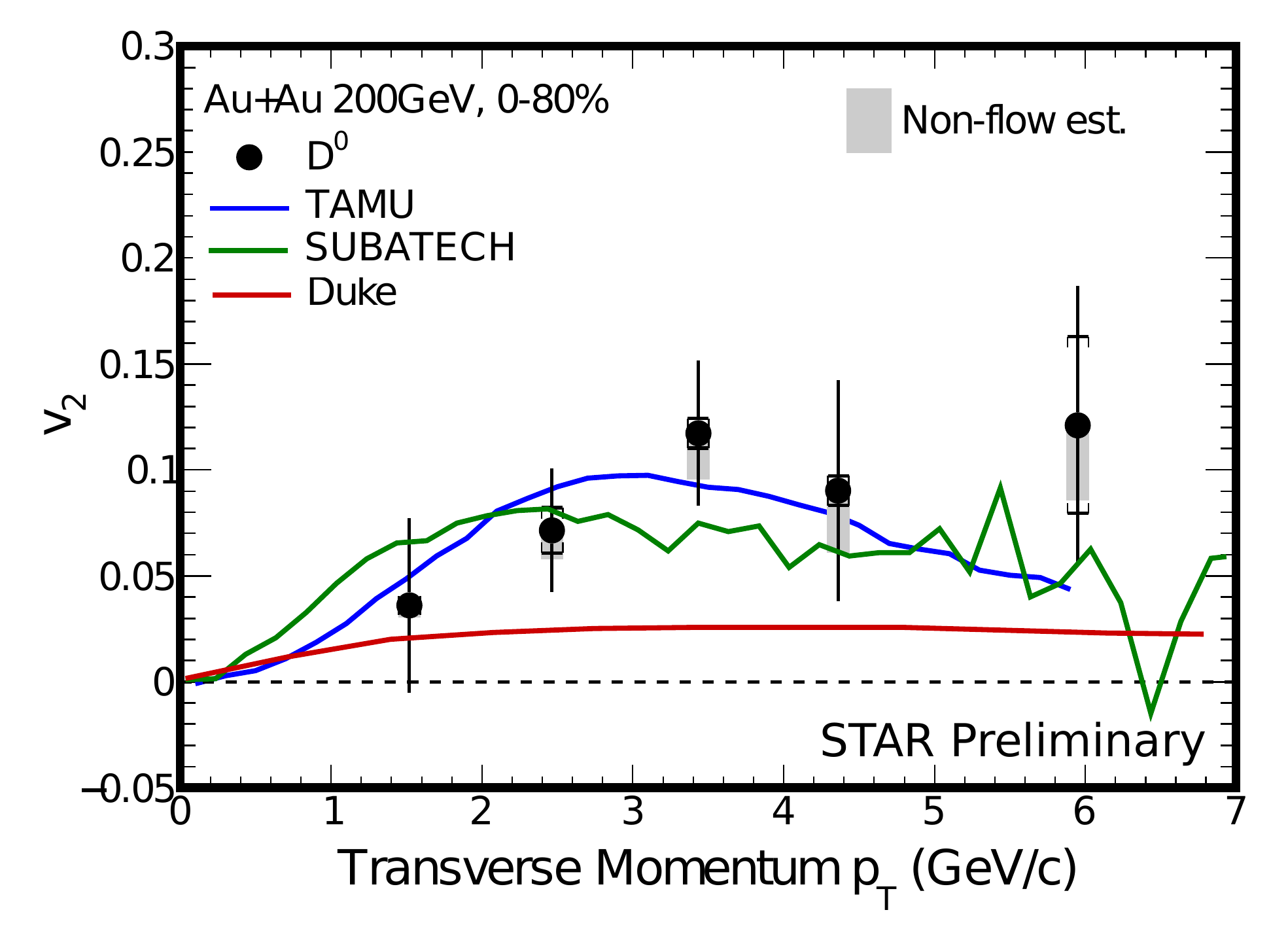}
    \includegraphics[width=0.33\textwidth]{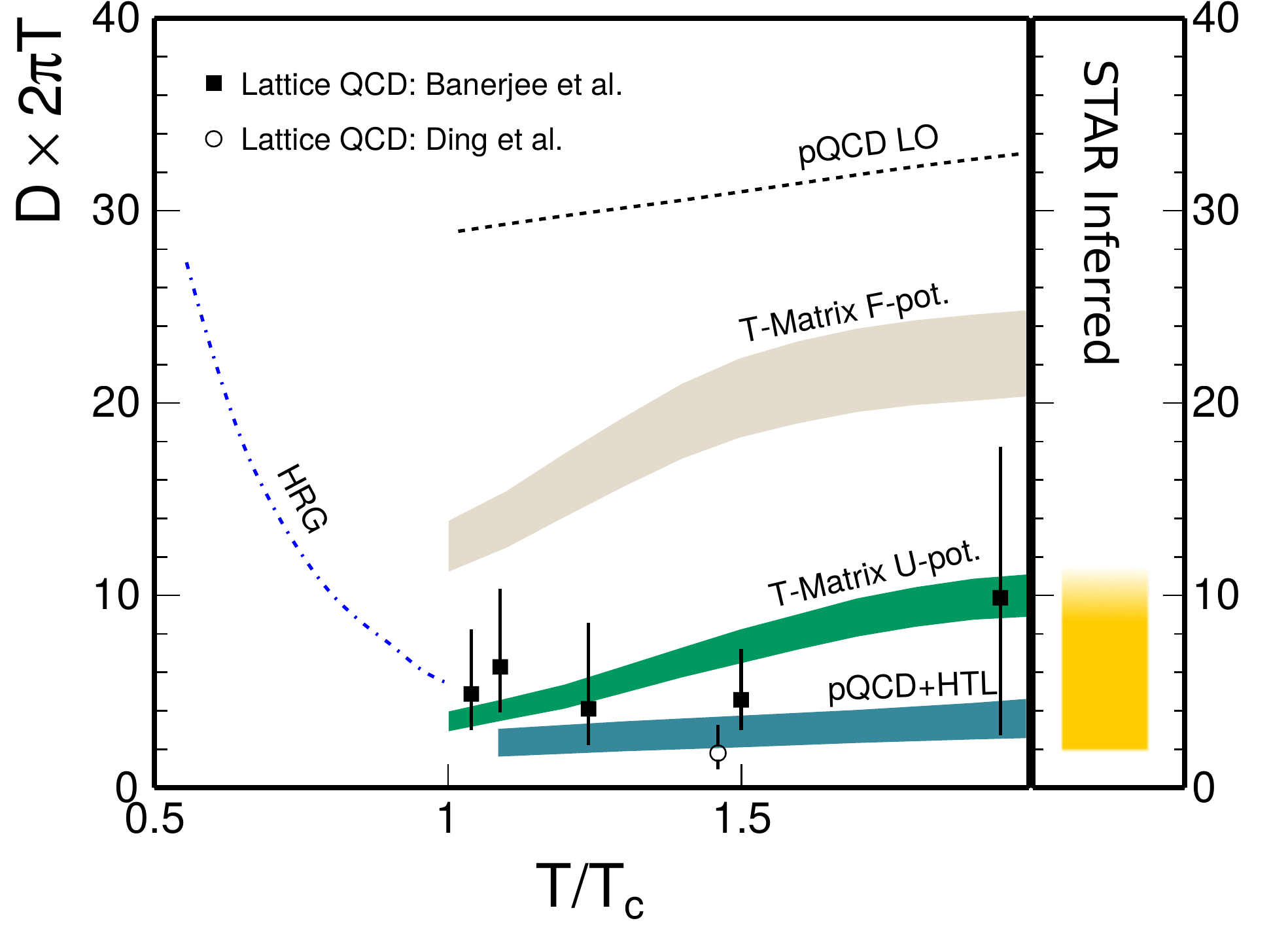}
    \caption{Comparison of $D^{0}$ nuclear modification factor (left) and $v_2$ (middle) to model calculations~\cite{rapp,subatech,duke}.
      (Right) Spatial diffusion coefficient of charm quarks ($T\ >\ T_c$) from various calculations and D mesons ($T\ <\ T_c$) in hadronic
      matter from Hadron Resonance Gas model, in units of the thermal wavelength~\cite{charmD,charmLQCD1,charmLQCD2,charmLQCD3,charmTMatrix1,charmTMatrix2,charmHTL,charmpQCD}.
    The right box shows the range of charm diffusion coefficient with which the models have an adequate description of STAR's $D^{0}$
  $R_{AA}$ and $v_{2}$ data.}

  \label{fig:d0_theory}
  \end{center}
\end{figure}

Figure \ref{fig:d0_theory} left and middle panels show a simultaneous comparison of $D^0$ $R_{AA}$ and $v_2$ to models that incorporate charm
diffusion in the medium~\cite{rapp,subatech,duke}. Apart from the small tension in the Duke model with the $v_2$ data, the models show a good description of STAR $D^0$ data,
The third panel shows calculations of charm spatial diffusion coefficient, as indicated
by the yellow band. STAR data are compatible with models that use $D\times 2\pi T$ of $2-\sim12$ demonstrating a strong charm medium interaction
that could possibly be at the quantum limit. The inferred range is also consistent with Lattice QCD calculations. While this comparison demonstrates the constraining
power of the data and the current status of theoretical models, more systematics studies are imperative to constrain the evolving bulk
parameters. Variations in these parameters have been shown to generate changes by a factor as large as $2$ in the heavy quark observables~\cite{gossiauxBulk}.

\label{sec:d0}

\subsection{Quarkonium production mechanisms and medium interactions}
\begin{figure}[h!]
  \begin{center}
    \includegraphics[width=0.35\textwidth]{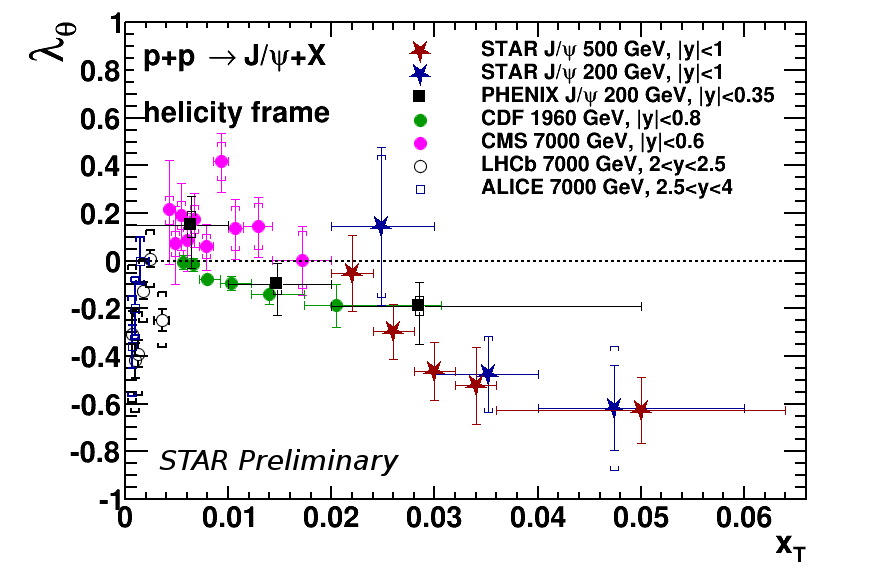}
    \includegraphics[width=0.27\textwidth]{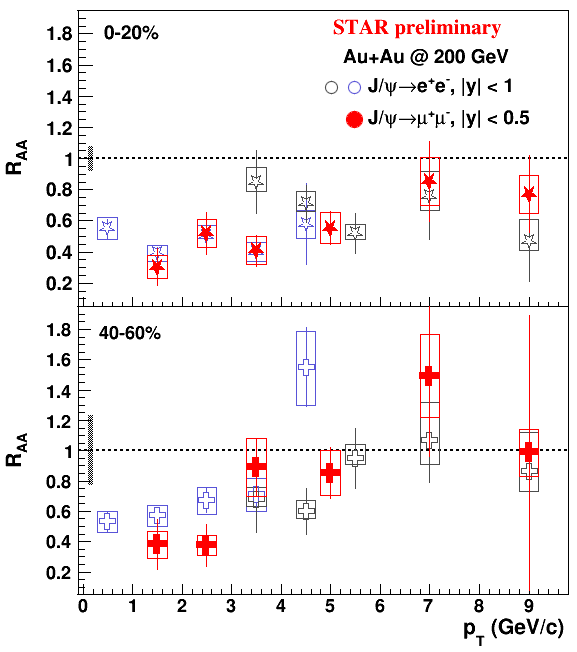}
    \includegraphics[width=0.35\textwidth]{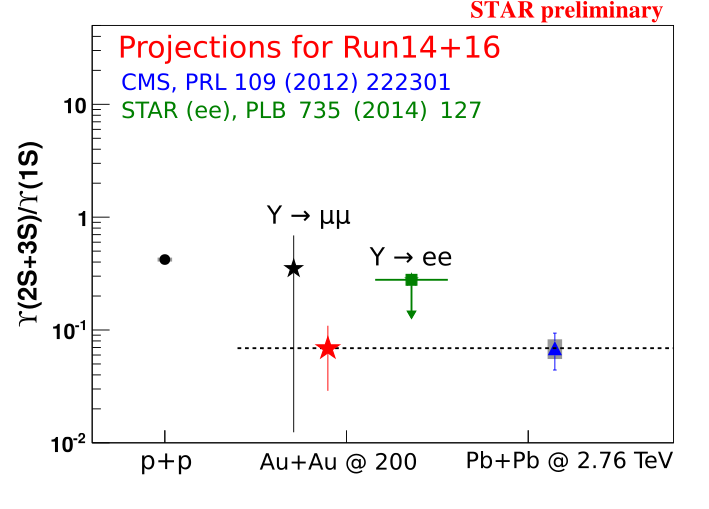}
    \caption{(Left) J/$\psi$ polarization parameter, $\lambda_{\theta}$, vs $x_T = 2p_T/\sqrt{s}$ in the helicity frame. (Middle) J/$\psi$ nuclear modification factor via di-electron (open symbols) and di-muon (filled symbols) channels as a function of $p_T$. (Right) Measurement of $\Upsilon(2S+3S)/\Upsilon(1S)$ in different collision systems at different energies~\cite{cmsUpsilon,starUpsilon}, also shown (red star) is
      the projection for measurement at STAR using 14.2 $nb^{-1}$ Au+Au data collected with the MTD in year 2014.}
  \label{fig:jpsi_upsilon}
  \end{center}
\end{figure}

There are several $J/\psi$ production models (eg: Color Octet Model, Color Evaporation Model, Non-Relativistic QCD) that correctly describe the production
yields in elementary particle collisions. However, the actual production mechanism is still unknown. STAR continues to carry differential measurements to
help distinguish between these models~\cite{barbara}.  Figure \ref{fig:jpsi_upsilon} left shows $J/\psi$ polarization parameter, $\lambda_{\theta}$, as 
a function of $x_T = 2p_T/\sqrt{s}$ in the helicity frame, at different collision energies. Common trend towards strong negative values with increasing $x_T$ values is observed.

Figure \ref{fig:jpsi_upsilon} middle shows the $R_{AA}$ of $J/\psi$ in Au+Au collisions at $\sqrt{s_{NN}} = 200$ GeV via the di-electron~\cite{jpsiSTAR} and the di-muon channels.
The suppression ($p_{T} < 5$ GeV/c) and the rising trend with $p_T$ are understood to be an interplay between dissociation and recombination effects at low $p_T$ and
the longer formation at high $p_T$ ~\cite{rongrong}. The MTD has been built to study and quantify the sequential melting of $\Upsilon$ states
in heavy-ion collisions. Figure \ref{fig:jpsi_upsilon} right shows the projection for the statistical uncertainty on $\Upsilon(2S+3S)/\Upsilon(1S)$ using the 14.2 $nb^{-1}$
which were sampled with the MTD in RHIC year 2014 run. A proof of principle analysis from 30\% of the collected statistics is shown as a black star on the figure.
The full data production is underway.

\subsection{Semi-inclusive charged jets in Au+Au collisions}
\begin{figure}[h!]
  \begin{center}
    \includegraphics[width=0.3\textwidth]{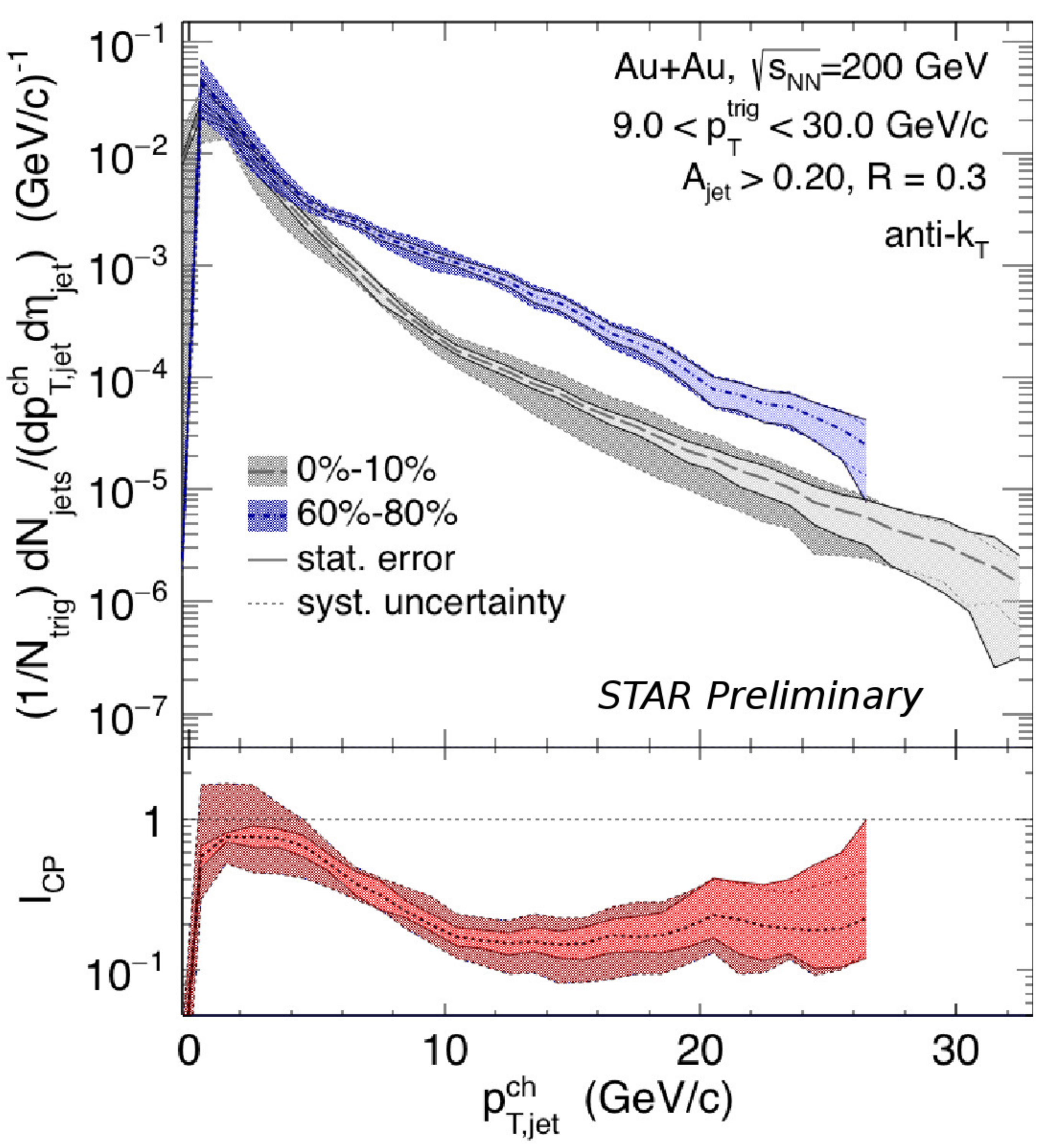}
    \includegraphics[width=0.3\textwidth]{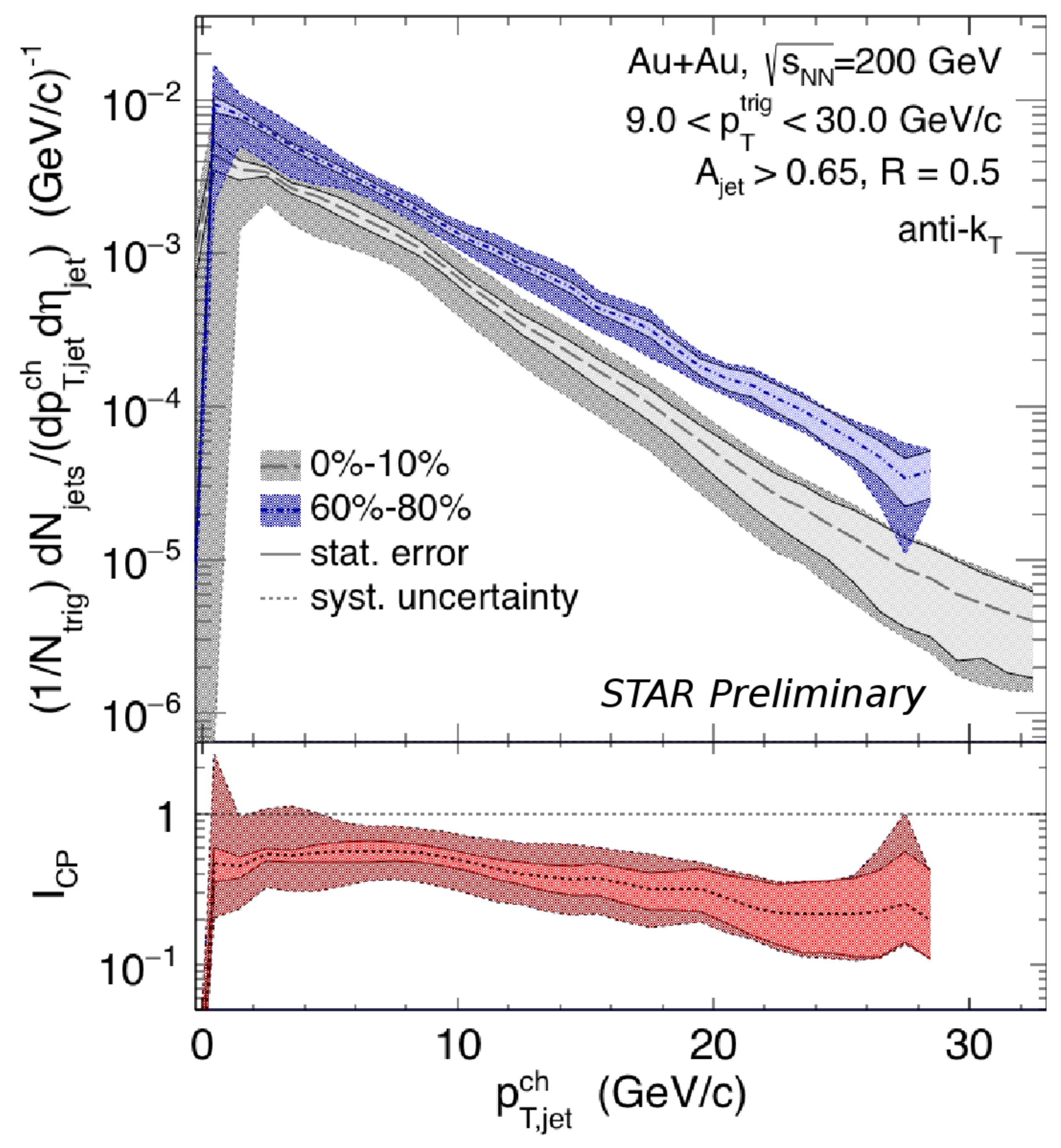}
    \caption{(Upper panels) Corrected recoil jet distributions for peripheral and central collisions.
    (Lower panels) $I_{CP}$, the ratio of central to peripheral yields. (Left) Cone size $R = 0.3$. (Right) Cone size $R = 0.5$.}
  \label{fig:jets}
  \end{center}
\end{figure}

Fully reconstructed jets are one of the main tools to constrain the medium transport parameters
and it is, thusly, an active area of research for current and planned future experiments. Full jet reconstruction in Au+Au collisions at $\sqrt{s_{NN}} = 200$ GeV have been carried out at
STAR using the anti-$k_{T}$ algorithm (R = 0.2 to 0.5) with a low IR-cutoff ($p_T > 0.2$ GeV/c)~\cite{peter}.
To address the challenge posed by the large soft and inhomogeneous underlying event background in measuring jet yields
STAR measures the semi-inclusive distributions of reconstructed charged particle jets recoiling from a high $p_{T}$ trigger hadron

\begin{equation}
  \frac{1}{N^h_{trig}}\frac{dN_{jet}}{dp_{T,jet}} = \frac{1}{\sigma^{AA\rightarrow h+X}}\frac{d\sigma^{AA\rightarrow h+jet+X}}{dp_{T,jet}}
\end{equation}

where $\sigma^{AA\rightarrow h+X}$ is the cross-section to generate a trigger hadron, $d\sigma^{AA\rightarrow h+jet+X}/dp_{T,jet}$ is the cross-section
for coincidence production of a trigger hadron and a recoil jet. A mixed-event technique has been developed
and used to treat the uncorrelated background. Figure \ref{fig:jets}, upper panels, show the fully corrected semi-inclusive recoil jet distributions
for central and peripheral events after unfolding $p_T$-smearing and instrumental effects. Results for cone radii R = 0.3 and 0.5 are shown for $9.0 < p_{T,trig} < 30.0$ GeV/c.
The ratios of recoil jet distributions in central to that in peripheral collisions, $I_{CP}$,  in lower panels show a clear suppression of yields in
central events for $p^{ch}_{T,jet} > 10$ GeV/c for both cone radii. The horizontal shifts in the range
where the $I_{CP}$ is flat ($10 < p^{ch}_{T,jet} < 20$) are $-6.3 \pm 0.6 \pm 0.8$ GeV/c and $-3.8 \pm 0.5 \pm 1.8$ GeV/c for R = 0.3 and R = 0.5, respectively.
The suppression and shift, and their reduction for larger cones might indicate an out-of-cone energy transport.

An open question is whether at jet-$Q^2$ scale the medium is effectively continuous or composed of quasi-particles. For this, STAR has also studied
the azimuthal distribution of jet axes with respect to the recoil trigger hadron in search for evidence of large angle scattering,
Moli\'ere scattering~\cite{peter}, whose presence would indicate hard scattering off quasi-particles.
The study shows so far no evidence of Mol\'ere scattering with the current statistical precision.

\section{Studies of chirality}

\subsection{Di-electrons}

\begin{figure}[h!]
  \begin{center}
    \includegraphics[width=0.4\textwidth]{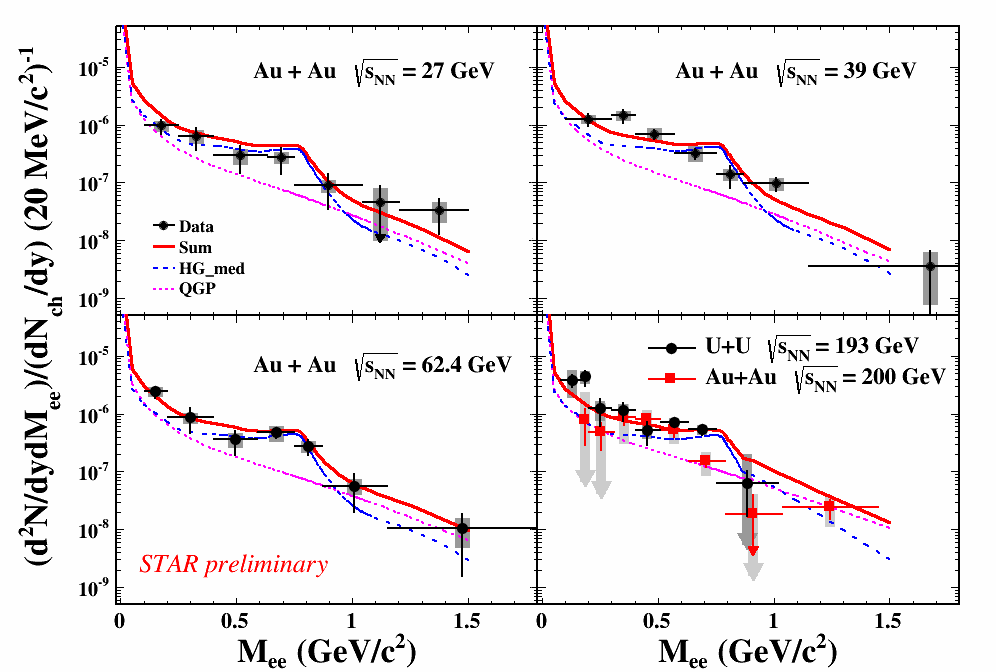}
    \includegraphics[width=0.4\textwidth]{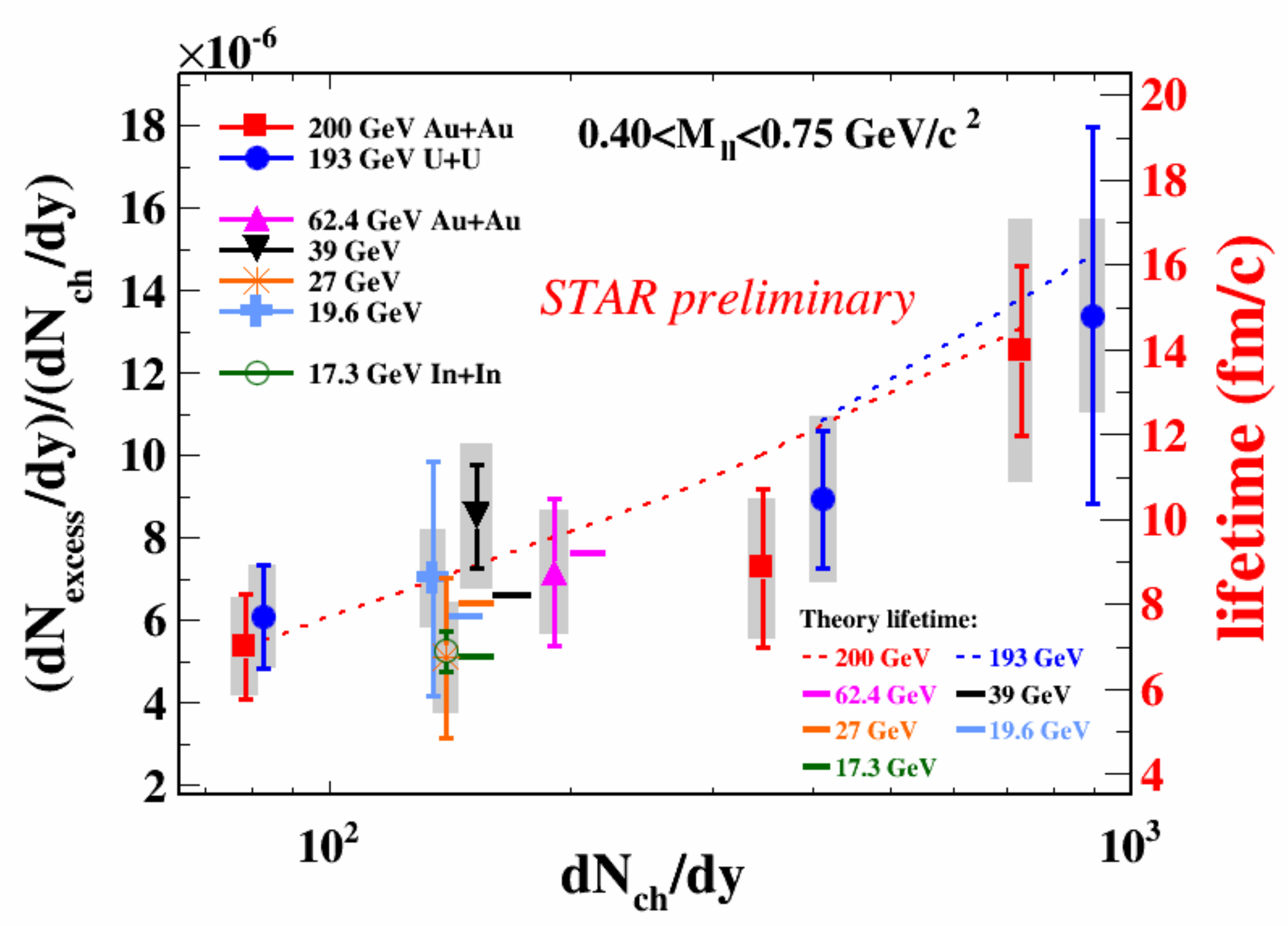}
    \caption{(Left) Acceptance-corrected di-electron excess invariant-mass spectra normalized by $dN_{ch}/dy$ in minimum-bias
      Au+Au collisions at $\sqrt{s_{NN}} = 27, 39, 62.4$ and $200$ GeV~\cite{starDielectrons} and U+U collisions at $\sqrt{s_{NN}} = 193$ GeV.
      Model calculations (red solid) with a broadened $\rho$ spectral function which is the sum of QGP radiation contribution and in hadronic
      medium broadening~\cite{rappDielectrons,heesDielectrons} is shown. (Right) Integrated di-electron excess invariant-mass spectra ($0.4 < M_{ll} < 0.75$ GeV/c)
      normalized by $dN_{ch}/dy$ in different collision species and at different collision energies~\cite{starDielectrons,na60Dielectrons} compared to
    the theoretical calculations of the fireball lifetimes~\cite{rappDielectrons}.}
  \label{fig:dielectrons}
  \end{center}
\end{figure}

The excess in the yields of di-leptons at low mass ($M_{ll} < 1$ GeV/$c^2$) has been long linked to chiral dynamics in heavy-ion collisions.
STAR has recently published its acceptance-corrected measurement of di-electron production in Au+Au collisions
at $\sqrt{s_{NN}} = 200$ and $19.6$ GeV~\cite{starDielectrons}. New measurement of di-electron production in U+U at $\sqrt{s_{NN}} = 193$ GeV,
a system with an expected 20\% increase in energy density~\cite{uuEnergy} and consequently a longer lifetime, have been presented at this conference~\cite{shuai}.
Figure \ref{fig:dielectrons} left shows the acceptance-corrected excess di-electron invariant-mass spectra normalized by $dN_{ch}/dy$ in minimum-bias
Au+Au collisions at $\sqrt{s_{NN}} = 27, 39, 62.4$ and $200$ GeV~\cite{starDielectrons} and U+U collisions at $\sqrt{s_{NN}} = 193$ GeV. Models with
collisional broadening of $\rho$-mesons~\cite{rappDielectrons,heesDielectrons} describe the data across all energies, centralities, and
$p_{T}$~\cite{shuai,starDielectrons}. Figure \ref{fig:dielectrons} right shows the integrated di-electron excess yields ($0.4 < M_{ll} < 0.75$ GeV/c$^2$)
normalized by and as function of $dN_{ch}/dy$. The yields show an increasing trend from peripheral to central collisions and from lower to higher energies,
which is predicted to be the case if the emission is correlated with the medium lifetime. The measurements are also compared to theoretical calculations
of medium lifetime~\cite{rappDielectrons}.

\subsection{Studies of chiral effects via PID correlations}
\begin{figure}[h!]
  \begin{center}
    \includegraphics[width=0.34\textwidth]{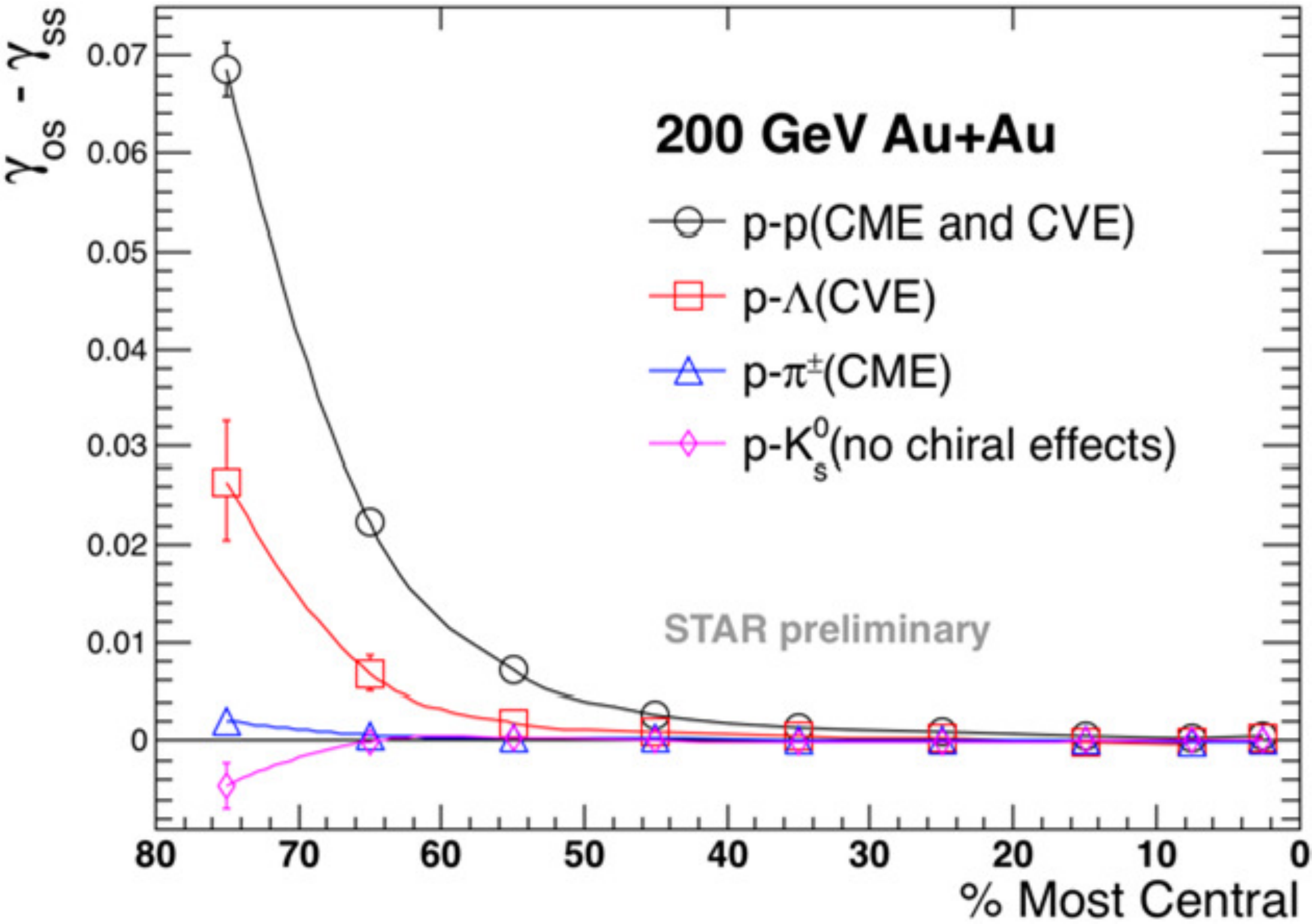}
    \includegraphics[width=0.27\textwidth]{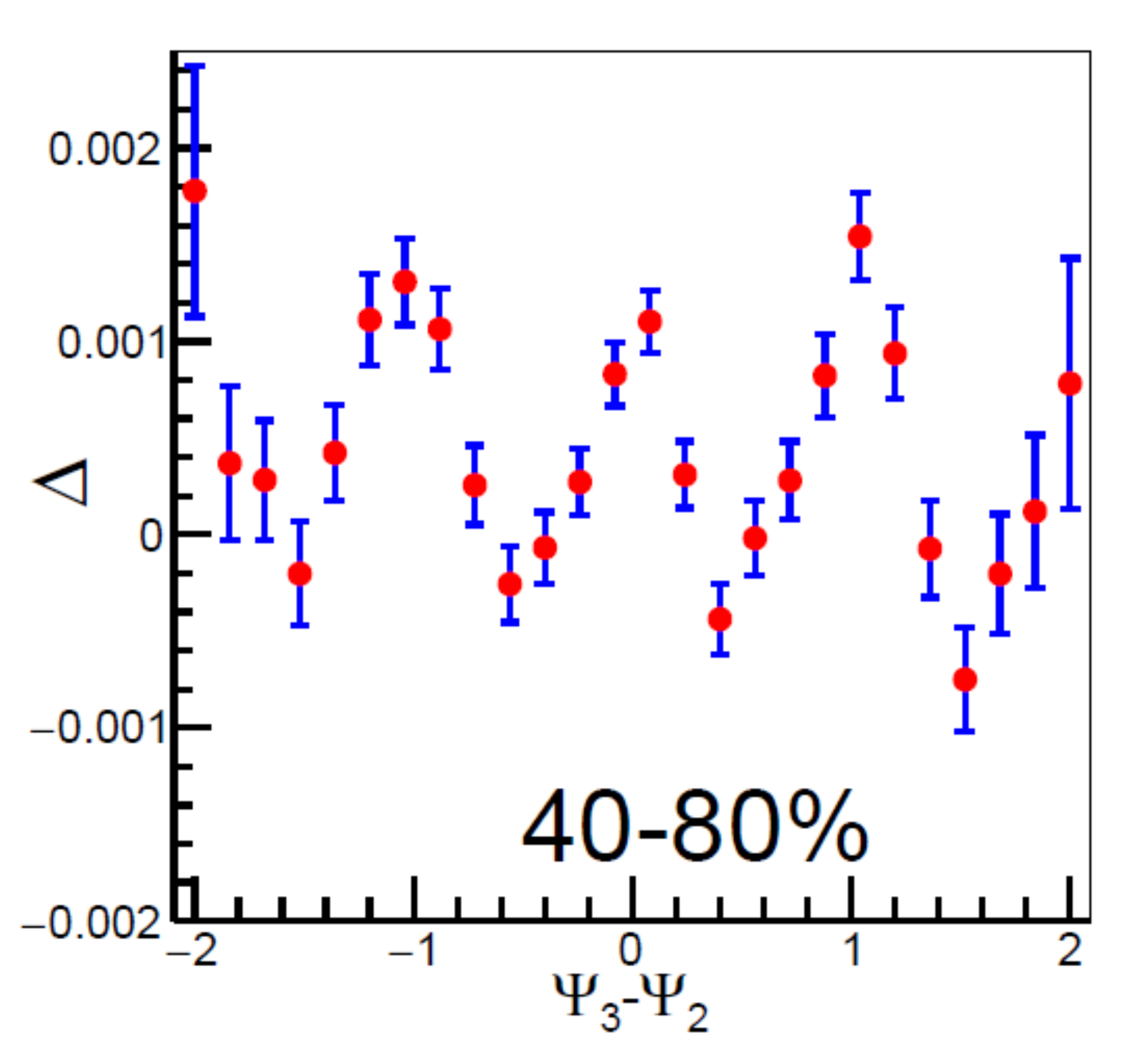}
    \caption{(Left) Opposite Sign (OS) - Same Sign (SS) three-point correlator of identified particles in Au+Au collisions at $\sqrt{s_{NN}} = 200$.
    (Right) charge separation parameter for non-central events with $v_2 \sim 0$.}
  \label{fig:cve}
  \end{center}
\end{figure}

The geometry of non-central collisions is expected to create strong magnetic fields at the early stages of heavy-ion collisions which if
coupled to a local chirality imbalance gives rise to electric charge separation, a phenomenon known as Chiral Magnetic Effect (CME). The same geometrical
configuration also creates an angular momentum that gives rise to fluid vorticity which, in turn, if coupled to local chirality imbalance creates
baryonic charge separation which is known as Chiral Vortical Effect (CVE). STAR uses a three point correlator ($\gamma = \left<\cos(\phi_{\alpha}+\phi_{\beta}-2\Psi_{RP})\right>$) to
search for signatures of CVE and CME. Correlations between different particle species have different sensitivities to these effects (see Fig. \ref{fig:cve} legend).
Figure \ref{fig:cve} left shows the difference of the correlator between opposite sign and same sign particles~\cite{liwen}. A clear hierarchy is observed, while this
hierarchical structure and its ordering meets the expectation of these chiral effects more studies of possible experimental backgrounds are needed and underway.
One such study has been presented at this conference~\cite{btu}. Figure\ref{fig:cve} right shows the charge separation parameter for
non-central events selected with observed $v_2 \sim 0$ where the $v_2$ related background is eliminated~\cite{cmwV2}. The charge separation is slightly positive on average,
but exhibits an oscillating dependence on the third harmonic plane relative to the second harmonic plane.
This suggests additional physics background related to the event shape and further investigations are underway.

\section{Studies of phase diagram structure}

\subsection{Onset of QGP formation and nature of phase transition}
\begin{figure}[h!]
  \begin{center}
    \includegraphics[width=0.7\textwidth]{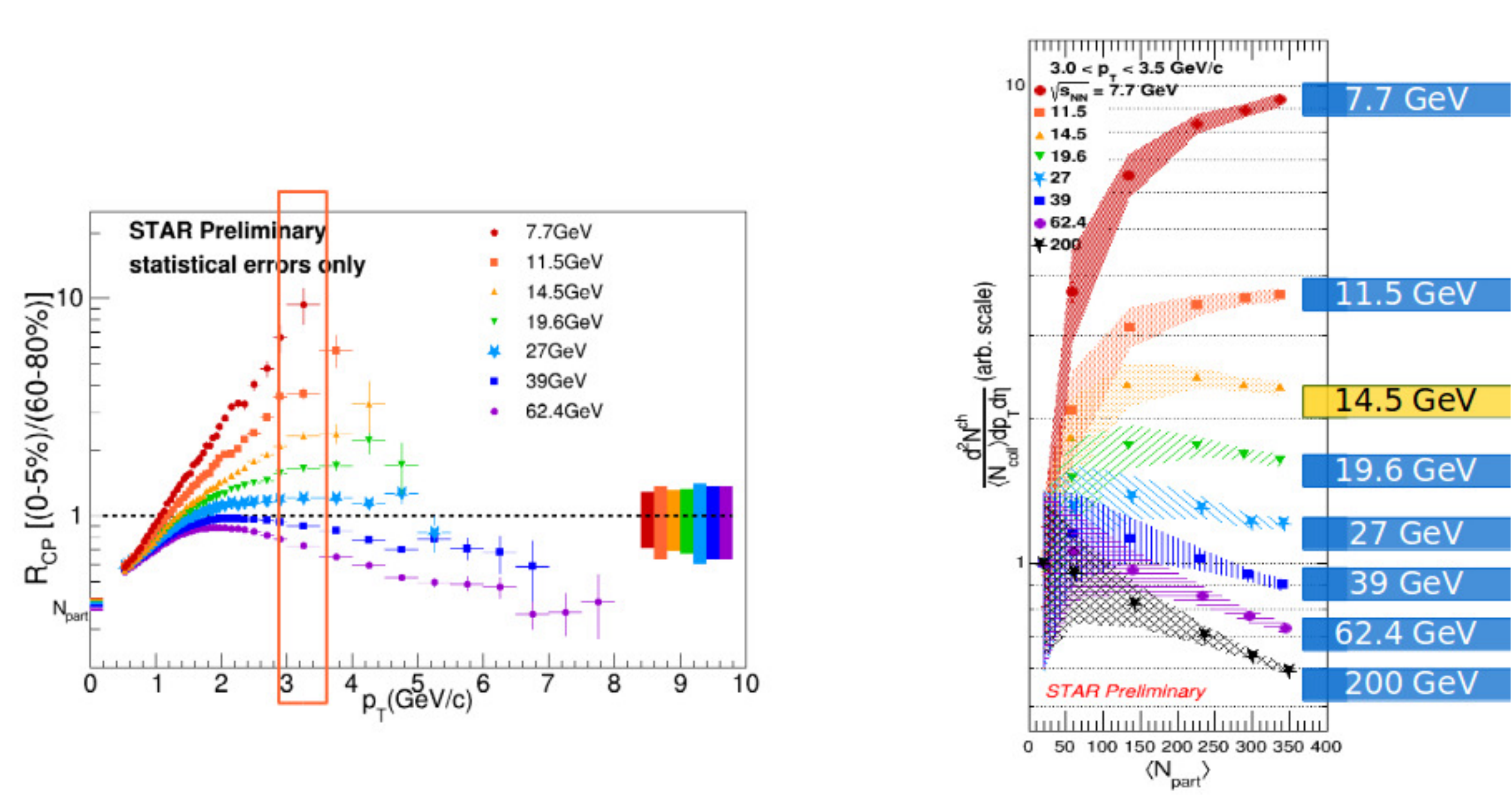}
    \caption{(Left) Charged hadron $R_{CP}$ in Au+Au collisions at different energies. (Right) Integrated yields ($ 3 < p_{T} < 3.5$ GeV) per number of binary
    collisions as a function of number of participants at different collision energies. The yields are normalized such that the yield in most peripheral bin is unity.}
  \label{fig:rcp}
  \end{center}
\end{figure}

High-$p_T$ hadron suppression is considered to be a key evidence of the formation of sQGP in heavy-ion collisions. It is only natural to study this observable in the BES to look for the onset
energy at which the strong suppression starts to appear. Figure \ref{fig:rcp} left shows charged hadron $R_{CP}$ at different BES-I energies~\cite{stephen}. $R_{CP}$ exhibits a smooth transition from strong
suppression at high energies to enhancement at lower beam energies. It is well known, however, that the so-called Cronin Effect, which gives rise to $p_{T}$ elongated production enhancement,
is more prominent at lower beam energies~\cite{cronin}, which complicates the interpretation of this observation. The complication can perhaps be resolved with a p+A BES to disentangle
the CNM enhancement and QGP suppression effects. The enhancement and suppression effects are proportional to centrality, so a study of
the centrality dependence of the charged hadron yields is expected to bring insight into the interplay of these effects. Figure \ref{fig:rcp} right shows
integrated yields ($3 < p_{T} < 3.5$ GeV) per number of binary collisions as a function of number of participants at different collision energies. The yields
are normalized such that the yield in most peripheral bin is unity. It is observed that at high beam energies suppression effects quickly dominate. At lowest beam
energies, however, the yields in most central collisions are highly enhanced compared to most peripheral collisions. At $\sqrt{s_{NN}} = 14.5$ GeV an approximate
flatness across centralities is observed, an apparent equilibration of enhancement and suppression effects at this energy and this $p_T$ region.
Theoretical calculations of the beam energy and impact parameter dependence could help understand the dynamics at play here and possibly uncover the suppression
effects from data.

\begin{figure}[h!]
  \begin{center}
    \includegraphics[width=0.4\textwidth]{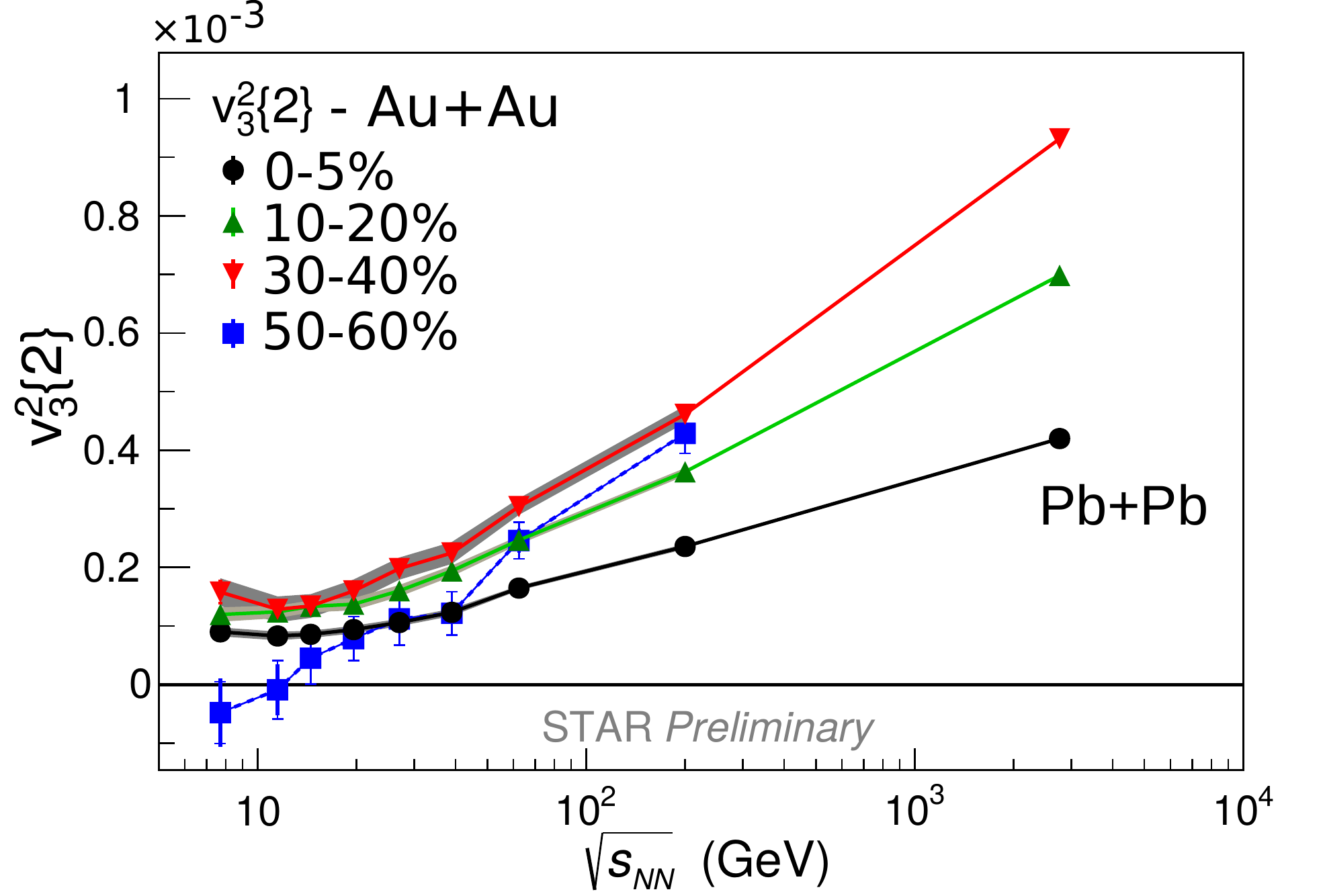}
    \includegraphics[width=0.4\textwidth]{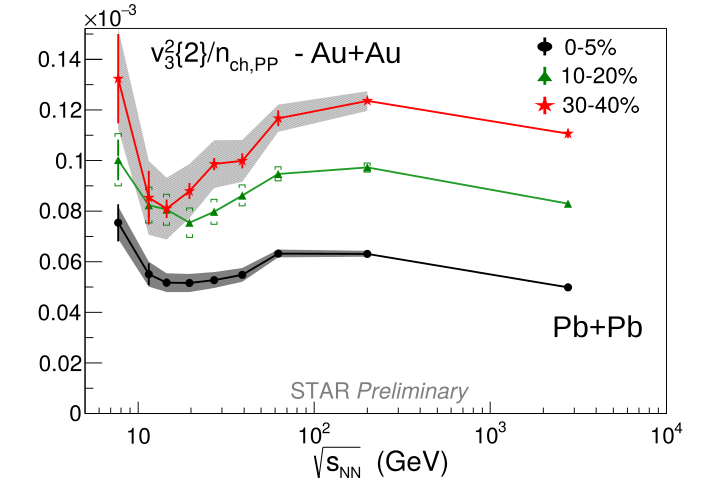}
    \caption{(Left) $v_{3}^2\{2\}$ in Au+Au collisions from STAR and in Pb+Pb collisions from ALICE~\cite{aliceV3} as a function of collision energy.
    (Right) $v_{3}^2\{2\}$ normalized by the multiplicity per participant pair ($n_{ch,PP} = dN_{ch}/dy/(N_{part}/2)$ as a function of collision energy.}
  \label{fig:v3}
  \end{center}
\end{figure}

Triangular flow is argued to be almost directly proportional to the duration of the low-viscosity phase in heavy ion collisions~\cite{v3Hannah}.
Figure \ref{fig:v3} left shows STAR measurements of charged hadron $v_3^2\{2\}$ in Au+Au collisions at $\sqrt{s_{NN}} = 7.7 - 200$ GeV ~\cite{liao} together with ALICE
measurement in Pb+Pb collisions at $\sqrt{s_{NN}} = 2.76$ TeV~\cite{aliceV3}. The data show that sizable $v_3$ in central to mid-central collisions
persists all the way down to the lowest beam energy inviting the question about the existence of a QGP phase even at those low energies. Peripheral collisions
$v_3$ is consistent with zero for $\sqrt{s_{NN}} \leq 14.5$ GeV. Figure \ref{fig:v3} right shows $v_3^2\{2\}$ scaled by the multiplicity per participant pair
($n_{ch,PP} = dN_{ch}/dy/(N_{part}/2)$) as a function of collision energy, an observable that is proportional to the system energy density. The data exhibit
a flat trend for energies $\sqrt{s_{NN}} = 7.7 - 20$ GeV in most central collisions, it is another open question whether this could be related to a softening of the Equation of State.

\begin{figure}[h!]
  \begin{center}
    \includegraphics[width=0.4\textwidth]{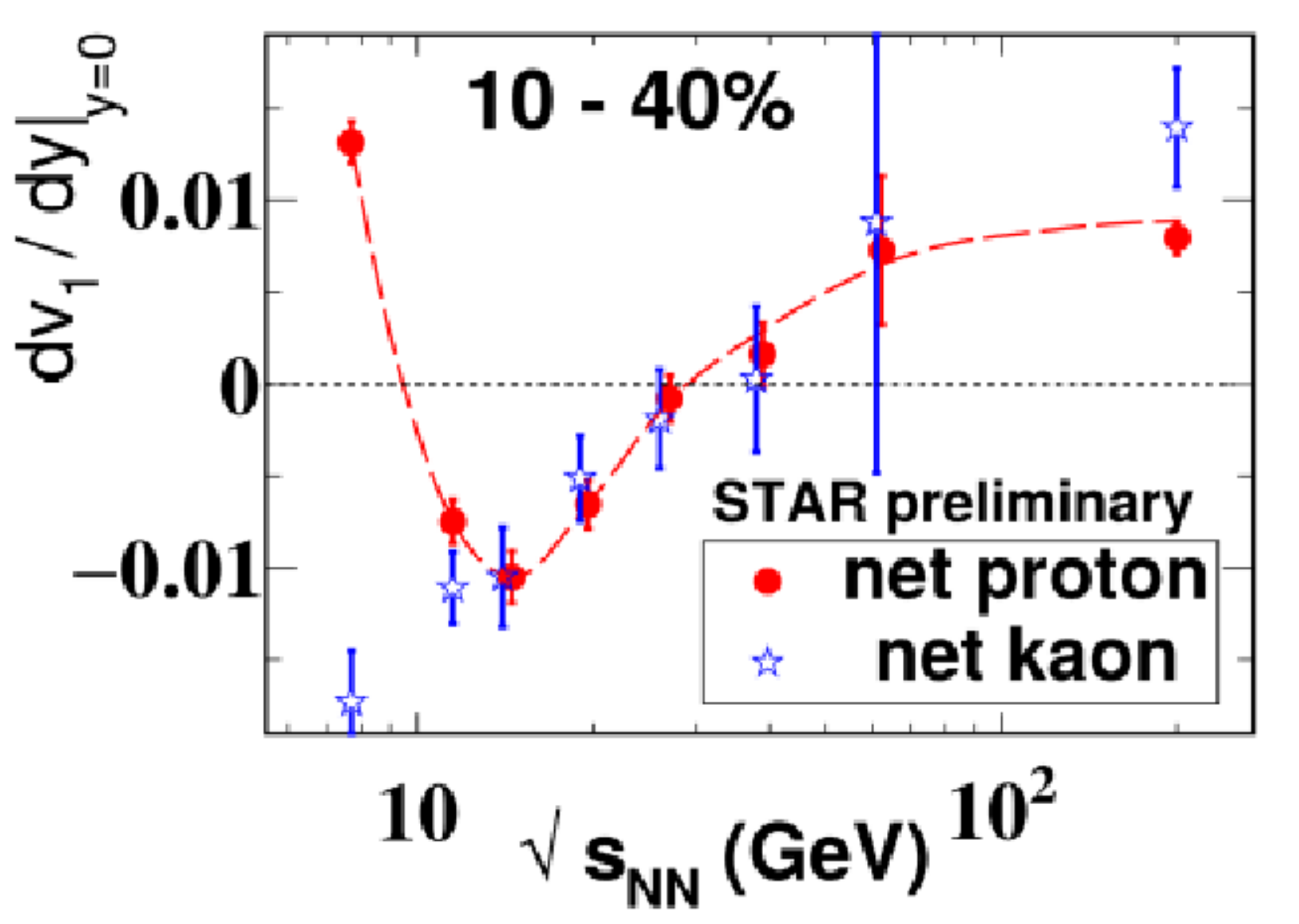}
    \includegraphics[width=0.46\textwidth]{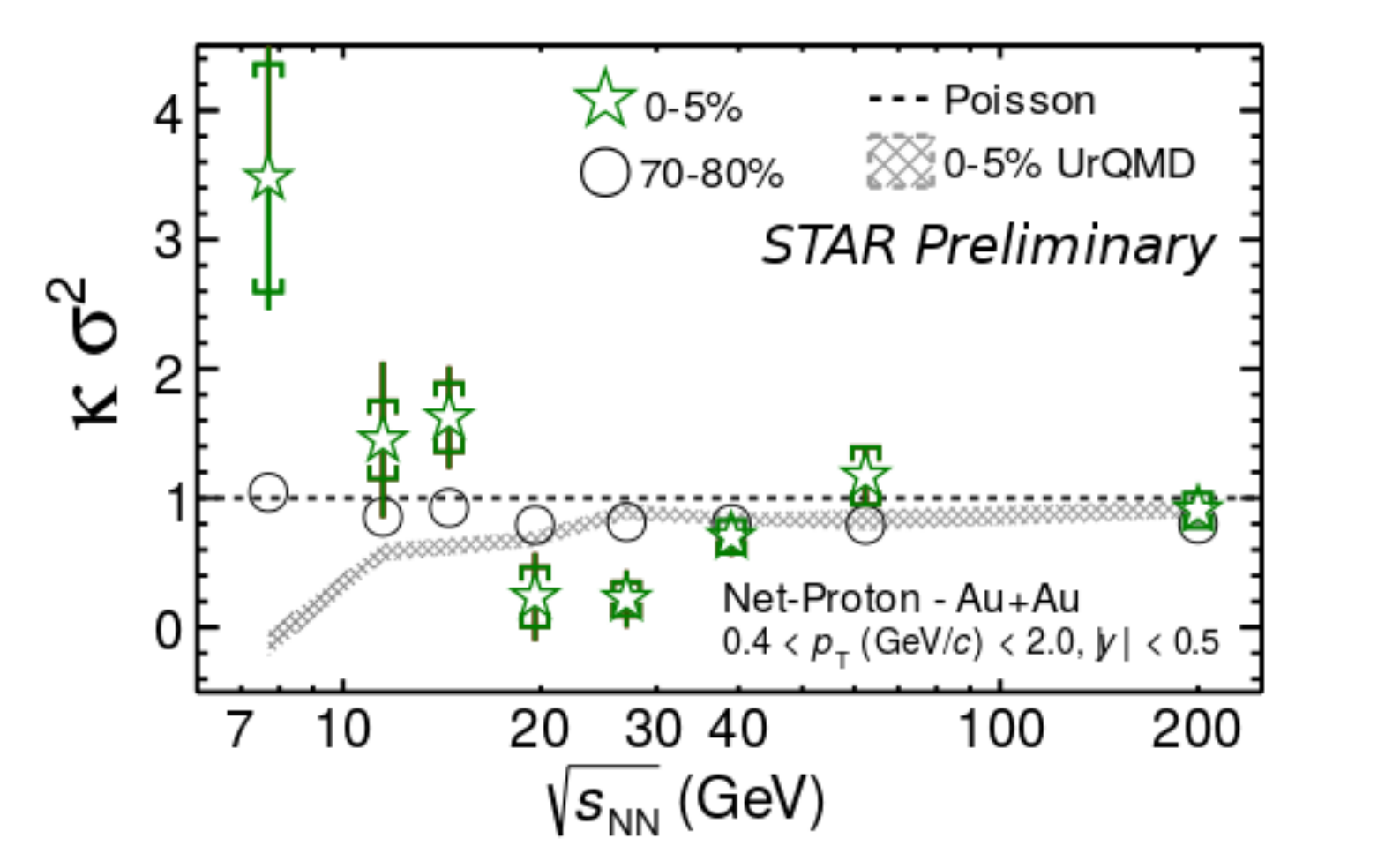}
    \caption{(Left) Net-proton and net-kaon $dv_{1}/dy|_{y=0}$ in Au+Au collisions at different collision energies. (Right) Net-protons $\kappa\sigma^2$ in Au+Au collisions as a function of collision energy compared to UrQMD calculations (hashed area) and Poisson predictions (dashed line) in 0-5\% (stars) and 70-80\% (circles) centralities.}
  \label{fig:v1}
  \end{center}
\end{figure}

The dip in STAR measurement of net-proton slope of directed flow at midrapidity~\cite{starV1}, $dv_1/dy|_{y=0}$, is due to an interplay between hydro, baryon
transport dynamics and baryon/anti-baryon annihilation,  so more understanding of these phenomena is needed as to whether a first order
phase transition is indicated by the data. So far all models failed to reproduce this dip structure. STAR has recently measured directed
flow of $p$, $\overline{p}$, $\Lambda$, $\overline{\Lambda}$, $K^{\pm}$, $K_S^0$ and $\pi^{\pm}$
in Au+Au collisions at the BES-I energies~\cite{prashanth}. Figure \ref{fig:v1} left shows net-proton and net-kaon $dv_1/dy|_{y=0}$. It is evident
that the data for net-kaon and net-proton coincide from $\sqrt{s_{NN}} = 200$ GeV all the way down to $14.5$ GeV and then starkly split at the lowest two energies.
More theoretical insight is needed to understand if this observation is related to the nature of the phase transition at these low energies.

\subsection{Search for the critical point}
Higher moments of conserved quantum numbers (charge, strangeness and baryon number) are expected to be sensitive to the proximity of the system
to a critical point (see~\cite{jochen} and references therein). It is also expected that higher order moments are more sensitive to criticality.
Figure \ref{fig:v3} right shows net-proton $\kappa\sigma^2$ in most central and peripheral collisions compared to UrQMD and Poisson expectations.
The measurement phase space has been recently extended from $0.4 < p_T < 0.8$ to $0.4 < p_T < 2.0$ GeV/c. The data in central collisions
clearly show a non-monotonic behavior deviating from the Poisson expectation of unity and that is absent from both peripheral and UrQMD expectations.
More work is being carried out to check if this experimental observable is sufficient to establish the existence of a critical point,
the role of the finite volume size in heavy-ion collisions and the random system evolution path in the phase diagram~\cite{jochen}.

It is currently agreed upon that more systematic studies of higher moments and their phase space evolution are needed from the experiment
side. This is one of the driving measurements behind RHIC plan for BES-II (2019-2020). STAR inner TPC upgrade will extend its coverage
from $|\eta| < 1$ to $|\eta| < 1.5$. In addition, an Event Plane Detector (EPD) upgrade at forward rapidities will provide more control on the measurement systematics
by giving a TPC-independent measurement of centrality and reaction plane.

\section{Summary and future outlook}

STAR has recently boosted its heavy flavor program with its high resolution Heavy Flavor Tracker (HFT) and Muon Telescope Detector (MTD).
$D^{0}$ $v_2$ measurement with the HFT indicates that charm flows at RHIC top energy. First results of quarkonium suppression from MTD have been shown
at this conference and more data on $\Upsilon$ states melting at RHIC top energy from full MTD datasets production are underway. STAR BES-I data exhibit non-monotonicity
in net-proton $\kappa\sigma^2$, baryon $dv_{1}/dy|_{y=0}$ and charged hadron scaled-$v_3^2\{2\}$ at the same energy range, trends that could
possibly be connected to the nature of phase transition at lower beam energies and the location of the critical point.


\bibliographystyle{elsarticle-num}

\begin{thebibliography}{00}
  \bibitem{giacomo} G. Contin, Nucl. Phys. A, these proceedings.
  \bibitem{Muller200584}B. M\"{u}ller, Nucl. Phys. A750 (2005) 84.
  \bibitem{rappHQReview} R. Rapp, H. van Hees, R. C. Hwa, X.-N. Wang (Ed.), Quark Gluon Plasma 4, World Scientific, p. 111 (2010).
  \bibitem{charmHTL}A. Andronic et al., arXiv:1506.03981.
  \bibitem{publishedD0}L. Adamczyk et al., STAR Collaboration, Phys. Rev. Lett. 113 (2014) 142301.
  \bibitem{pionsRaa}B.I. Abelev et al., STAR Collaboration, Phys. Lett. B655 (2007) 104.
  \bibitem{ksV2}B.I. Abelev et al., STAR Collaboration, Phys. Rev. C77 (2008) 54901.
  \bibitem{rapp} M. He, R. J. Fries, R. Rapp, Phys. Rev. C86 (2012) 014903, and private communications.
  \bibitem{guannan} G. Xie, for the STAR Collaboration, Nucl. Phys. A, these proceedings.
  \bibitem{michael} M. Lomnitz, for the STAR Collaboration, Nucl. Phys. A, these proceedings.
  \bibitem{subatech} M. Nahrgang, J. Aichelin, S. Bass, P. B. Gossiaux, K. Werner, Phys. Rev. C91 (2015) 014904, and private communications.
  \bibitem{duke} S. Cao, G.-Y. Qin, S. A. Bass, Phys. Rev. C88 (2013) 044907, and private communication.
  \bibitem{gossiauxBulk} P. B. Gossiaux, S. Vogel, H. van Hees, J. Aichelin, R. Rapp, M. He, M. Bluhm, arXiv:1102.1114.
  \bibitem{charmD}M. He, R. J. Fries, R. Rapp, Phys. Rev. Lett. 110 (2013) 112301.
  \bibitem{charmLQCD1}H. Ding, A. Francis, O. Kaczmarek, F. Karsch, H. Satz, et al., Phys. Rev. D86 (2012) 014509.
  \bibitem{charmLQCD2}D. Banerjee, S. Datta, R. Gavai, and P. Majumdar, Phys. Rev. D85 (2012) 014510.
  \bibitem{charmLQCD3}O. Kaczmarek, Nucl. Phys. A931 (2014) 633.
  \bibitem{charmTMatrix1}F. Riek and R. Rapp, Phys. Rev. C82 (2010) 035201.
  \bibitem{charmTMatrix2}K. Huggins and R. Rapp, Nucl. Phys. A896 (2012) 24.
  \bibitem{charmpQCD}B. Svetitsky, Phys. Rev. D37 (1988) 2484.
  \bibitem{rongrong} R. Ma, for the STAR Collaboration, Nucl. Phys. A, these proceedings.
  \bibitem{cmsUpsilon}S. Chatrchyan et al., CMS Collaboration, Phys. Rev. Lett. 109 (2012) 222301.
  \bibitem{starUpsilon}L. Adamczyk et al., STAR Collaboration, Phys. Lett. B735 (2014) 127.
  \bibitem{shuai} S. Yang, for the STAR Collaboration, Nucl. Phys. A, these proceedings.
  \bibitem{uuEnergy} D. Kiko\l{}a, G. Odyniec, R. Vogt, Phys. Rev. C84 (2011) 054907.
  \bibitem{starDielectrons} L. Adamczyk et al., STAR Collaboration, Phys. Lett. B750 (2015) 64 and Phys. Rev. C92 (2015) 024912.
  \bibitem{rappDielectrons} R. Rapp, Phys. Rev. C63 (2001) 054907 and R. Rapp, H. van Hees, arXiv:1411.4612.
  \bibitem{heesDielectrons} H. van Hees, R. Rapp, Nucl. Phys. A806 (2008) 229.
  \bibitem{na60Dielectrons} R. Arnaldi et al., NA60 Collaboration, Eur. Phys. J. C59 (2009) 607.
  \bibitem{cronin} I. Vitev, M. Gyulassy, Phys. Rev. Lett. 89 (2002) 252301.
  \bibitem{aliceV3} K. Aamodt et al., ALICE Collaboration, Phys. Rev. Lett. 107 (2011) 032301.
  \bibitem{peter} P. Jacobs, A. Schmah, for the STAR Collaboration, Nucl. Phys. A, these proceedings.
  \bibitem{barbara} B. Trzeciak, for the STAR Collaboration, Nucl. Phys. A, these proceedings.
  \bibitem{jpsiSTAR1} L. Adamczyk et al., STAR Collaboration, Phys. Lett. B722 (2013) 55 and Phys. Rev. C90 (2014) 024906.
  \bibitem{liwen} L. Wen, for the STAR collaboration, poster (948) at QM15.
  \bibitem{btu} B. Tu, N. N. Ajitanand, for the STAR collaboration, poster (123) at QM15.
  \bibitem{cmwV2} L. Adamczyk et al., STAR Collaboration, Phys. Rev. C89 (2014) 044908.
  \bibitem{stephen} S. Horvat, for the STAR Collaboration, Nucl. Phys. A, these proceedings.
  \bibitem{v3Hannah} J. Aunvine, H. Petersen, Phys. Rev. C88 (2013) 064908.
  \bibitem{liao} L. Song, for the STAR Collaboration, Nucl. Phys. A, these proceedings.
  \bibitem{starV1} L. Adamczyk, STAR Collaboration, Phys. Rev. Lett. 112 (2014) 162301.
  \bibitem{prashanth} P. Shanmuganathan, for the STAR Collaboration,  Nucl. Phys. A, these proceedings.
  \bibitem{jochen} J. Th\"{a}der, for the STAR Collaboration, Nucl. Phys. A, these proceedings.
\end{thebibliography}

\end{document}